\documentclass[11pt]{amsart}
\usepackage{geometry}                
\usepackage{graphicx}
\usepackage{amssymb}
\usepackage{epstopdf}
\usepackage{array}
\usepackage{subfigure}
\usepackage{url}
\usepackage{hyperref}
\usepackage{xspace}
\usepackage{lscape}
\usepackage{isorot}

\usepackage{color}

\definecolor{gray}{rgb}{0.7,0.7,0.7}

\usepackage{supertabular}
\usepackage{hhline}
\newcommand\arraybslash{\let\\\@arraycr}
\makeatother
\title[Slime mould tactile sensor]{Slime mould tactile sensor}
\author[Adamatzky]{Andrew Adamatzky\\ University of the West of England, Bristol, UK}
\address[Adamatzky]{University of the West of England, Bristol, United Kingdom}

\begin{document}

\maketitle

\begin{abstract}

Slime mould \emph{P. polycephalum} is a single cells visible by unaided eye. The cells shows a wide spectrum of intelligent behaviour. 
By interpreting the behaviour in terms of computation one can make a slime mould based computing device. The Physarum computers are 
capable to solve a range of tasks of computational geometry, optimisation and logic. Physarum computers designed so far 
lack of localised inputs. Commonly used inputs --- illumination and chemo-attractants and -repellents --- usually act on 
extended domains of the slime mould's body.  Aiming to design massive-parallel tactile inputs for slime mould computers we 
analyse a temporal dynamic of \emph{P. polycephalum}'s electrical response to tactile stimulation. In experimental laboratory 
studies we discover how the  Physarum responds to application and removal of a local mechanical pressure with 
electrical potential impulses  and changes in its electrical potential oscillation patterns. 

\vspace{0.5cm}

\end{abstract}

\section{Introduction}

Tactile sensors are quint-essential components of modern robotic devices. Most robots, especially those built for medical applications,
rely on a vital information from their tactile sensors when physically interacting with their environment, human operators and 
subjects~\cite{Lucarotti_2013, Rocha_2008, Hamed_2008, Mukai_2008, Muhammad_2011,Lucarotti_2013,Cutkosky_2008}. 
Novel and original  implementations of tactile sensors include  
\begin{itemize}
\item arrays of piezo-electric polymers with auxiliaries, converting force into voltage~\cite{Dahiya_2009},
\item arrays of electro-active polymers~\cite{wang_2008}, 
\item polymer hair cell sensors~\cite{Engel_2006a}, 
\item force sensitive conductive rubber~\cite{Ohmukai_2012}, 
\item elastomers filled with carbon nanotubes~\cite{Engel_2006}, 
\item POSFET tactile arrays~\cite{Dahiya_2009}, 
\item pressure sensitive conductive rubber~\cite{Kato_2008}, and  
\item ionic polymer metal composites~\cite{wang_2009}. 
\end{itemize}

With a rise of bio-inspired and hybrid wetware-hardware 
robots~\cite{frasca_2004, floreano_2009, adamatzky_komosinksi, prescott_2012, gruber_2012}  interest in technological 
developments drifted away from solid materials to a soft matter~\cite{Tiwana_2012}. Successful implementations of bio-inspired 
and soft sensors include 
\begin{itemize}
\item a bio-mimetic sensor, which employs a conductive fluid encapsulated in elastic container and uses
deformation of the elastic container in transduction~\cite{Wettels_2008}, 
\item flexible capacitive micro-fluidic based sensors~\cite{Wong_2012},
\item patterns of micro-channels filled with eutectic gallium-indium~\cite{Park_2010, Park_2011}, 
\item live cell sensors \cite{Taniguchi_2010}, and  
\item  bio-hybrid sensors encapsulating living fibroblasts as a part of transduction system~\cite{Cheneler_2012}.
\end{itemize}

In a series of previous works, see overview in~\cite{adamatzky_physarummachines},  we developed a concept and fabricated experimental laboratory 
prototypes of amorphous bio-computing devices --- Physarum machines.  A Physarum machine is a programmable amorphous biological computing device experimentally implemented in plasmodium of \emph{P. polycephalum}.  \emph{Physarum polycephalum} belongs to the species of 
order \emph{Physarales},  subclass \emph{Myxogastromycetidae}, class \emph{Myxomycetes}, division \emph{Myxostelida}. It is commonly known as a true, acellular or multi-headed slime mould. Plasmodium is a `vegetative' phase, a single cell with a myriad of diploid nuclei.  The plasmodium is visible to the unaided eye. The plasmodium looks like an amorphous yellowish mass with networks of protoplasmic tubes. The plasmodium behaves and moves as a giant amoeba. It feeds on bacteria, spores and other microbial creatures and micro-particles~\cite{stephenson_2000}.  The plasmodium's foraging behaviour can be  interpreted as a computation:  data are represented by spatial distribution of attractants and repellents, and  results are represented by a structure of Physarum's protoplasmic  network. In such specification a plasmodium can solve computational problems with natural parallelism, including optimisation on graphs, computational geometry, logic and robot control, 
see details in~\cite{adamatzky_physarummachines}.

A Physarum machine is programmed by configurations of repelling and attracting gradients: chemical substances, temperature and illumination. 
These quantities are often difficult to localise, which makes a precise, fine-grained, input of spatial data into Physarum machines problematic.
A tactile input of information could be a solution. Thus in present we evaluate a feasibility of Physarum to act as a tranducer:  
to transform a tactile stimulation or a mechanical pressure to a distinctive pattern of an electrical activity. 
We study how parameters of the oscillations change in response to an application and removal of a solid light-weight insulators to 
 Physarum's protoplasmic tubes or sheet-shaped parts.

\section{Methods}

\begin{figure}[!tbp]
\centering
\subfigure[]{\includegraphics[width=0.8\textwidth]{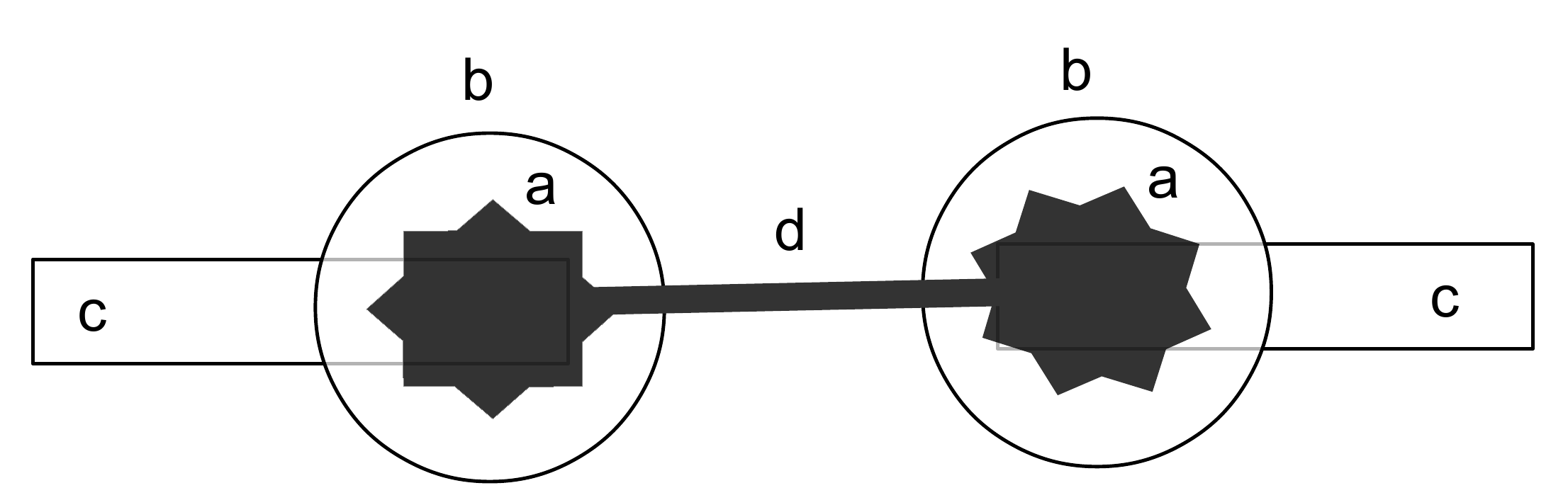}}
\caption{A scheme of experimental setup: (a)~Physarum, (b)~agar blobs, (c)~electrodes, (d)~protoplasmic tube. All parts of Physarum 
shown in dark grey form a single cell.}
\label{scheme}
\end{figure}

Plasmodium of \emph{Physarum polycephalum} was cultivated in plastic lunch boxes (with few holes punched in their lids for ventilation) on wet kitchen towels and fed with oat flakes. Culture was periodically replanted to a fresh substrate. Electrical activity of plasmodium was recorded with  
ADC-24 High Resolution Data Logger  (Pico Technology, UK).  A scheme of experimental setup is shown in Fig.~\ref{scheme}. Two blobs of agar 
2 ml each (Fig.~\ref{scheme}b) were placed on electrodes (Fig.~\ref{scheme}c) stuck  to a bottom of a plastic Petri dish (9~cm). Distance between proximal sites of electrodes is always 10~mm. Physarum was inoculated on one agar blob. We waited till Physarum colonised the first blob, where it was inoculated, and propagated towards and colonised the second blob. When second blob is colonised, two blobs of agar, both colonised by Physarum (Fig.~\ref{scheme}a), became connected by a single protoplasmic tube  (Fig.~\ref{scheme}d). We discounted experiments more than one tube was formed between the blobs because because patterns of oscillation were affected by interactions between potential waves travelling along interlinked protoplasmic tubes.

Loads  were applied either to a protoplasmic tube  (Fig.~\ref{scheme}d) connecting the blobs or to a sheet of Physarum covering agar blob on recording electrode. The following events of tactile stimulation were studied in the experiments.
\begin{itemize}
\item TR($w$): A piece of  glass capillary weighting $w$~g, $w=0.01, 0.05, 0.1, 0.15$, is applied across a protoplasmic tube (Fig.~\ref{scheme}d) connecting two blobs of agar colonised by Physarum, 
\item TA($w$): a piece of glass capillary weighting $w$~g is lifted (c. 10~min after application) of a protoplasmic tube; 
\item BA($w$): a plastic load $w$~g, $w=0.05, 0.2, 0.35, 3$ is applied to a Physarum colonising agar blob on a recording electrode, 
weights are 0.5-2~mm thick plastic discs, 5-7~mm in diameter, and BlueTak ball weighting  3~g,
\item BR($w$) --- a load $w$~g is lifted (c. 10~min after application) from Physarum colonising agar blob.  
\end{itemize}

\section{Results}

\begin{figure}[!tbp]
\centering
\includegraphics[width=1.1\textwidth]{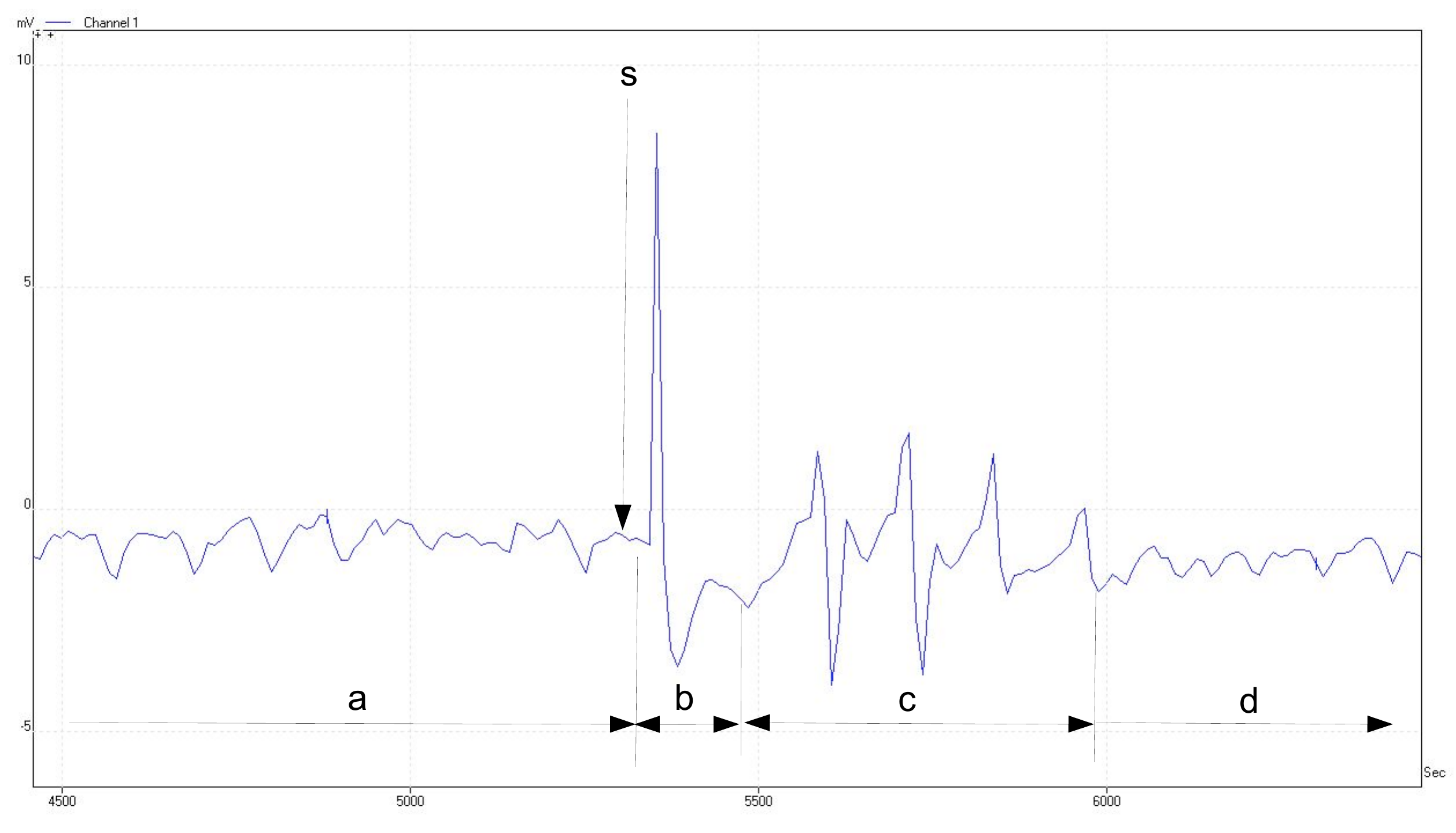}
\caption{Physarum's response to TA(0.01) event.  Vertical axis is an electrical potential value in mV, horizontal axis is time in seconds.
Glass capillary weighting 0.01~g is applied across protoplasmic tube 
at 5370~sec from beginning of recording. (a)~Oscillator activity before stimulation. (b)~Immediate response to stimulation. 
(c)~Prolonged response to stimulation. (d)~Return to normal oscillatory activity.}
\label{stagesscheme}
\end{figure}

Undisturbed Physarum exhibits periodic changes, or oscillations, of its surface electrical potential. 
A typical normal oscillation of a surface potential has amplitude of 0.1 to  5 mV, could be less subject to location of electrodes, and 
period  1-4~min~\cite{iwamura_1949, kamiya_1950,kashimoto_1958}. Exact pattern  of electric potential oscillations  depends 
on a physiological state and age of Physarum culture and particulars of experimental setups~\cite{achenbach_1980}. 
In 1939 Heilbrunn and Daugherty discovered that peristaltic activity of protoplasmic tubes is governed by oscillations of electrical 
potential propagating along the tubes~\cite{heilbrunn_1939}. An exact nature of correlation between electrical and contractile 
oscillation of plasmodium is still unclear, there is a view that these two oscillations are governed by the same mechanism but 
may occur independently on each other~\cite{simons_1981}.

A typical response of Physarum towards application of a load is shown in Fig.~\ref{stagesscheme}. Physarum exhibits more or less 
classical oscillations before stimulation (Fig.~\ref{stagesscheme}a), shape of oscillatory waves is a bit distorted, possibly due to 
minor branches of the tube connecting the blobs and electrodes  (Fig.~\ref{scheme}d).  A segment of a glass capillary is placed 
across protoplasmic tube 
(Fig.~\ref{scheme}d) at 5370th sec from the beginning of recording (Fig.~\ref{stagesscheme}s). Physarum demonstrates two types 
of responses to application of this load:  an immediate response with a high-amplitude impulse (Fig.~\ref{stagesscheme}b) and
a prolonged response with changes in oscillation pattern (Fig.~\ref{stagesscheme}c). The immediate response is a 
high-amplitude spike: its amplitude is 12.33~mV and its duration is 150~sec. The prolonged response is an envelop of increased 
amplitude oscillations.  An average amplitude of oscillations before stimulation, in the example shown in Fig.~\ref{stagesscheme},  
was 2.3~mV and duration  of each wave was 120~sec. The amplitude of waves in the prolonged response became 5.29~mV
 with a duration of a wave slightly increased to 124~sec.  

\begin{figure}[!tbp]
\centering
\includegraphics[width=1.2\textwidth]{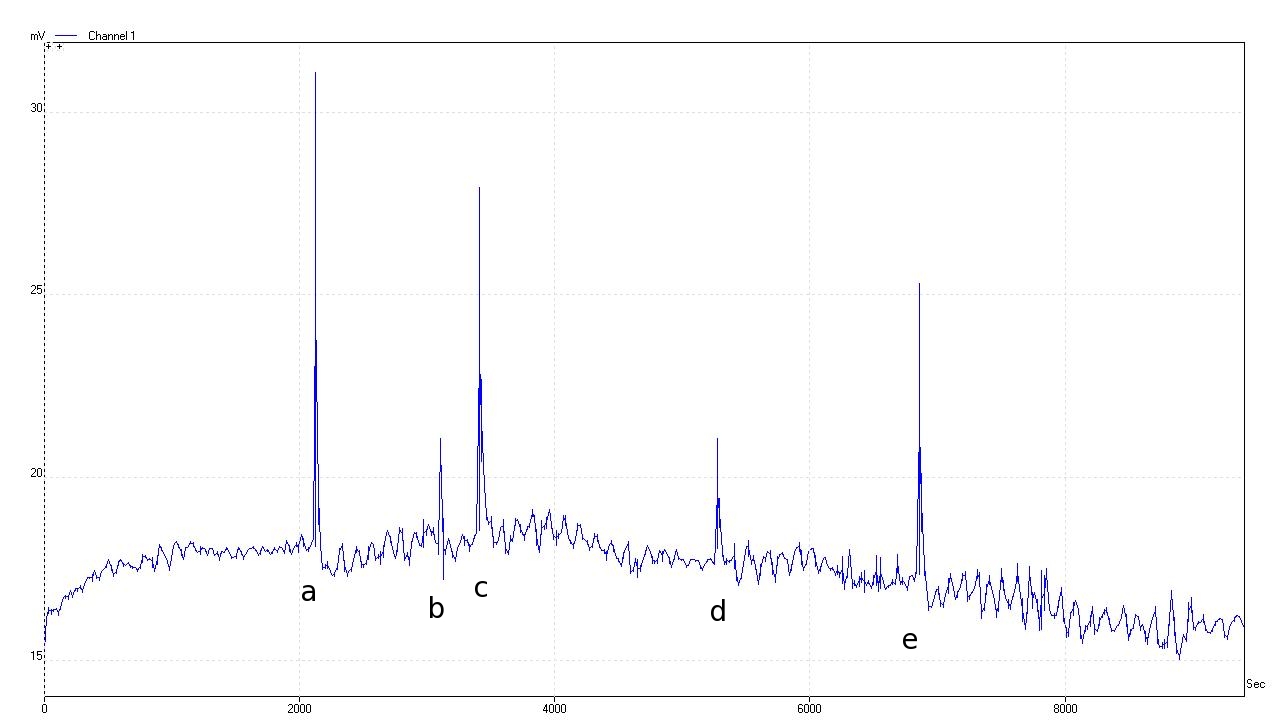}
\caption{An exemplar scenario of the Physarum tactile response. 
Vertical axis is an electrical potential value in mV, horizontal axis is time in seconds.
(a)~TA(0.05) at 2150~sec. 
(b)~Protoplasmic tube partially collapsed but restored its connectivity in c. 5~min.
(c)~TR(0.05) at 3430~sec.
(d)~TA(0.05) at 5300~sec.
(e)~TR(0.05) at 6870~sec.}
\label{weight_001_005_01_exp050301}
\end{figure}

An example of Physarum responses to repeated cycle of applying and removing loads --- events TA(0.05) and TR(0.05) --- is shown in Fig.~\ref{weight_001_005_01_exp050301}. A segment of capillary tube, 0.05~g, was applied and removed twice. Physarum reacted to first application with a spike of amplitude 11.5~mV and duration 121~sec. Next minor spike of 3~mV (marked 'b' in Fig.~\ref{weight_001_005_01_exp050301}) was due to the protoplasmic tube partially collapsing.   A response impulse to removal of the load has amplitude 9.7~mV and duration 158~sec (Fig.~\ref{weight_001_005_01_exp050301}c).  In this example, an impulse response to a second application of 0.05~g load was a bit less pronounced: c. 3.6 mV amplitude and 71~sec duration  (Fig.~\ref{weight_001_005_01_exp050301}d). Response to second removal of the load was as strong as the response to first removal: 8.1~mV amplitude and 135~sec duration (Fig.~\ref{weight_001_005_01_exp050301}e). 

\begin{table}[!tbp]
\caption{Characteristics of a high-amplitude impulse response, calculated in 21 experiments. 
$V$  is an average amplitude of Physarum electrical potential response,  in mV, W 
is an average duration of the response, in sec. 
$\sigma(\cdot)$ is a standard deviation. SNR($\cdot$) is a signal to noise ratio calculated for amplitude and width of the impulse. }
\begin{tabular}{l|llllll}
		& $V$ & $\sigma(V)$ & $W$ &  $\sigma(W)$  & SNR($V$) & SNR($W)$ \\ \hline 
TA(0.01) 	& 5.8 &  7.0 	  & 120 &  106.1 & 12.9 & 1.3\\
TR(0.01) 	& 8.1 &  10.2  & 93 &  33.9 & 9.5	&1.1\\
TA(0.05) 	& 8.1 &  10.0 & 68.3 &  20.2   & 20.3	& 0.8 \\
TR(0.05)	& 11.4 &  20.1 & 94.5 &  89.7 & 24.8 & 1.4\\
TA(0.1)   	& 1.7 &  0.6   & 100.5 &  19.5  & 9.4 & 1.5\\
TR(0.1) 	& 3.0 &  0.2 & 144.5 &  113.5 & 8.6 & 2.1\\
TA(0.15) 	& 2.0 &  0.8   & 79.3 &  32.3    & 4.8 & 1.6\\
TR(0.15)	& 2.2 &  1.1 & 69.0 &  46.0 & 7.9 & 0.8 \\
BA(0.2)   	& 4.2 &  5.0   & 203.3 &  177.7 & 42 & 1\\
BR(0.2)	&3.5 &  2.7 & 72.0 &  38.7 & 31.8 & 0.4\\
BA(0.35) 	& 2.1 &  0.7   & 216.0 &  144.0 & 5.4 & 1.7 \\ 
BR(0.35)	&2.9 &  1.1 & 325.5 &  105.5 & 10 & 2.9 \\
BA(3)      	& 0.7 &  0.3  & 157.0 &  88.8    & 3.5 & 1.8 \\
BR(3)		&0.9 &  0.4 & 156.5 &  44.5 &4.8 & 1.8\\
 \end{tabular}
 \label{statisticsimpulse}
\end{table}

\begin{figure}[!tbp]
\centering
\includegraphics[width=0.9\textwidth]{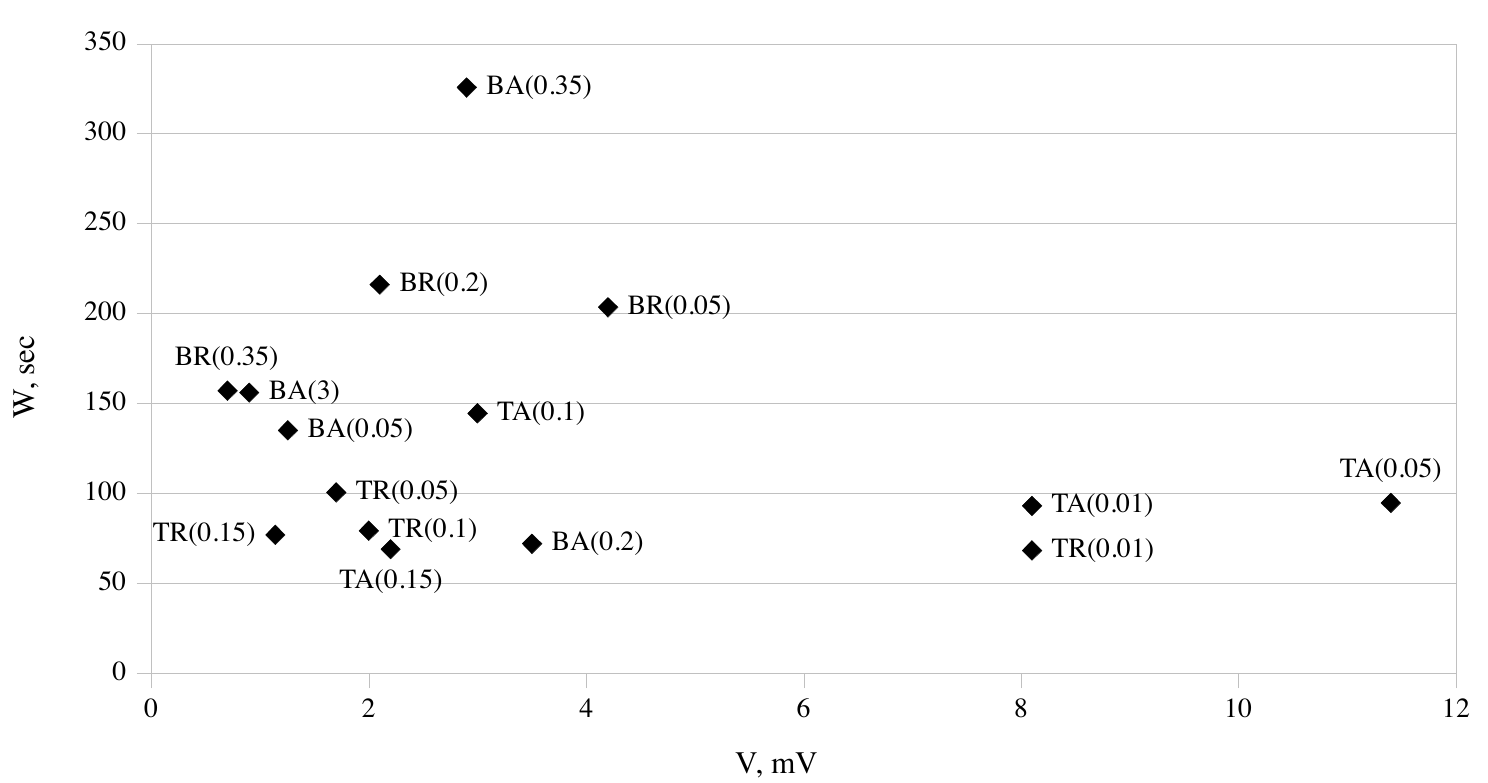}
\caption{Average amplitude $V$, in mV, versus average duration $W$, in sec, of a
high-amplitude impulse response of Physarum to application  
(TA($w$), BA($w$)) and removal (TR($w$), BR($w$))  of  $w$~g load to a protoplasmic tube   
(TA($w$), TR($w$)) and agar blob (BA($w$), BR($w$)).  }
\label{impulseplot}
\end{figure}

Parameters of the high-amplitude impulse  calculated in 21 experiments, are shown in Tab.~\ref{statisticsimpulse}
and Fig.~\ref{impulseplot}. Large values of standard deviation is the first thing one could notice 
in Tab.~\ref{statisticsimpulse}, 2nd and 4th columns. Both amplitude and width of an immediate response spike are distributed over
 a large interval of values. This is particularly visible in amplitude response, where in some cases deviation exceeds average values, e.g. 
 events TA(0.01), TR(0.01), TA(0.05), TR(0.05) and BA(0.2). In overall, amplitude of a response to application and removal of a load to a protoplasmic tube varies much more than amplitude of a response to application and removal of a load to agar blob colonised by Physarum. Deviation of the impulse width from its average value is much lower, in average 55\% (Tab.~\ref{statisticsimpulse}, 4th column). 
 
 We did not observe any statistically conclusive correlation between weights of loads applied or removed and amplitude or width of the immediate response impulse. Physarum responds with a wider impulse when loads are applied to or removed of an agar blob colonised by Physarum than
 when a protoplasmic tube is mechanically stimulated (Fig.~\ref{impulseplot}). This may be because larger areas of Physarum are affected when agar 
 blobs are stimulated. Amplitude of an immediate response to stimulation of a protoplasmic tube is, in overall, higher than amplitude of the response to stimulation of Physarum on agar blobs, e.g. see plots corresponding to T(0.05), T(0.01) and TR(0.01) in Fig.~\ref{impulseplot}.

\begin{table}[!tbp]
\caption{Characteristics of electrical potential oscillations recorded before ($V'$, $W'$) and after ($V''$, $W''$) application (TA($w$), BA($w$)) and 
removal  (TR($w$), BR($w$)) of $w$~g load to a protoplasmic tube (TR($w$), TA($w$)) and an agar blob occupied by Physarum 
(BR($w$), BA($w$)), calculated in 21 experiment. $V'$ and $V''$ are average amplitudes of oscillations before ($V'$) and after ($V''$) stimulation, 
$W'$ and $W''$ are average durations of a single impulse in oscillator pattern before ($W'$) and after ($W''$) stimulation; $\sigma(\cdot)$ is a standard deviation.}
\begin{tabular}{l|llllllllll}
	&	$V'$ &  $\sigma(V')$	&	$W'$ &  $\sigma(W')$	&	$V''$ &  $\sigma(V'')$	&	$W''$ &  $\sigma(W'')$	&	$V''/V'$	&	$W''/W'$	\\ \hline
TA(0.01)	&	0.45 &  0.27	&	91.2 &  31.1	&	0.74 &  0.52	&	81.67 &  34.09	&	1.64	&	0.9	\\
TR(0.01)	&	0.85 &  0.26	&	84.5 &  25.84	&	1.7 &  0.65	&	99.25 &  17.0	&	2	&	1.17	\\
TA(0.05)	&	0.4 &  0.12	&	89.6 &  22.43	&	0.64 &  0.3	&	112 &  14.11	&	1.6	&	1.25	\\
TR(0.05)	&	0.46 &  0.26	&	69.0 &  31.16	&	0.77 &  0.59	&	107.75 &  13.38	&	1.67	&	1.56	\\
TA(0.1)	&	0.18 &  0.2	&	66 &  41.01	&	0.31 &  013	&	49.50 &  40.31	&	1.72	&	0.75	\\
TR(0.1)	&	0.35 &  0.07	&	54.5 &  47.38	&	0.23 &  0.04	&	89.5 &  45.96	&	0.66	&	1.64	\\
TA(0.15)	&	0.42 &  0.38	&	48.5 &  51.62	&	0.61 &  0.52	&	84 &  9.9	&	1.45	&	1.73	\\
TR(0.15)	&	0.28 &  0.32	&	81.5 &  4.95	&	0.67 &  0.56	&	103 &  52.33	&	2.42	&	1.26	\\
BA(0.2)   	&  0.1 &  0.07  	& 204 &  47.3   &	0.16 &  0.74  & 129 &  32.1  &  1.6		& 0.63	\\
BR(0.2) 	& 0.11 &  0. 04  	&  192 &  34.2  &	 0.21 &  0.16  & 174 &  16.5 &  1.91	& 0.91	\\ 
BA(0.35) 	& 0.39 &  0.21  	& 129 &   32.7 &	 0.63 &  0.34 & 136.7 &  23.8 &  1.61	& 1.06	\\ 
BR(0.35) 	&  0.4 &  0.12  	& 112 &  29.8        &	 	0.66 &  0.27  & 106.5 &  21.9 &  1.65	& 0.95	\\ 
BA(3)      	&  0.29 &  0.08 	& 86 &   13.5       &	 	0.32 &  0.18 &  98 &  9.4	&  1.1		& 1.14 \\
BR(3)      	&  0.19 &  0.03	&  88 &   16.2       &	 	0.39 &  0.21 &  75 &  18.3 &  2.07	& 0.85	\\
\end{tabular}
\label{statisticsoscillationtube}
\end{table}
				
Experimental data on prolonged response to stimulation are provided in Tab.~\ref{statisticsoscillationtube}. Physarum increases amplitude of oscillations in response to application and removal of  almost all loads studied. Most pronounced amplitude of a potential impulse is observed
in stimulation events TR(0.15), BR(3), TR(0.01), BR(0.2). Width of oscillation impulses increases after mechanical stimulation 
in eight of fourteen stimulation events. In such situations a width of oscillations after stimulation is  f1.06, BA(0.35), to  1.73, TA(0.15), times wider than width of oscillations before stimulation (Tab.~\ref{statisticsoscillationtube}). No increase in the width is observed in events TA(0.01), TA(0.1), 
BA(0.2), BR(0.2), BR(0.35), BR(3).

\begin{figure}[!tbp]
\centering
\subfigure[]{\includegraphics[width=0.9\textwidth]{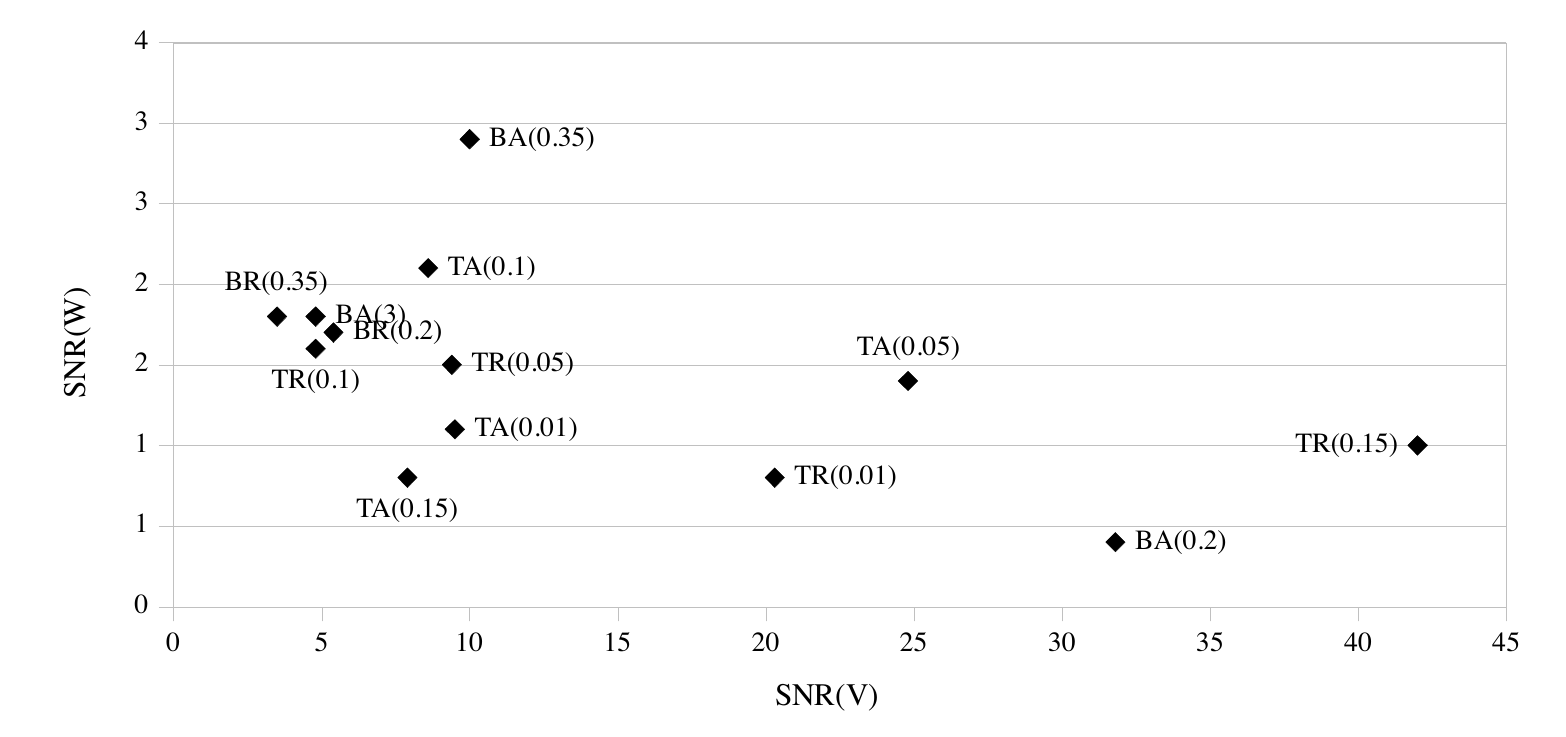}}
\subfigure[]{\includegraphics[width=0.9\textwidth]{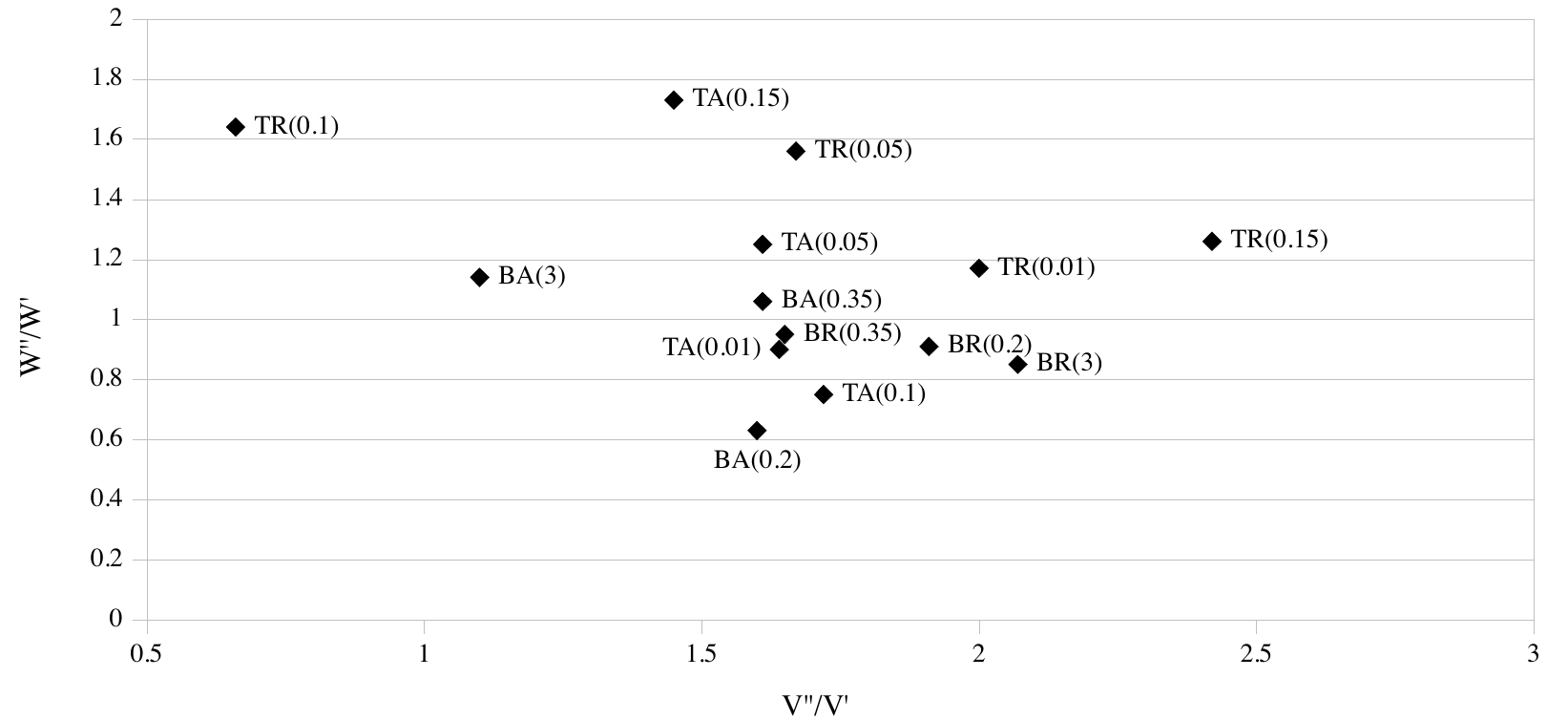}}
\caption{
(a) SNR of amplitude SNR($V)$ versus SNR of duration SNR($W$) of high-amplitude impulse 
response to mechanical stimulation.
(b) Mechanical stimulation induced changes in oscillations patterns: $V'/V''$ versus $W'/W''$ map. 
$V'$ and $V''$ are average amplitudes of oscillations before and after stimulation, 
$W'$ and $W''$ are average widths of oscillations before and after stimulation.  }
\label{SNRall}
\end{figure}

What could be a signal to noise ratio (SNR) of Physarum tactile sensors?  To calculate SNR we can consider normal oscillations of
surface electrical potential is a `noise' and immediate response impulse is a `signal'. Thus SNR of amplitude, SNR($V$), 
is a ratio of an average amplitude of the immediate response impulse $V$ to average amplitude $V'$ of oscillations before
 stimulation. SNR of width,  SNR($W$), is  $W/W'$, where $W$ is an average width of immediate response impulse and 
 $W'$ is an average width of oscillatory impulses before stimulation. 

SNRs of immediate response to various stimulation events are shown in Fig.~\ref{SNRall}a, see exact values in 
Tab.~\ref{statisticsimpulse}.  Highest values of SNR($V$), over 20, are recorded in experiments when loads 0.05~g and 
0.2~g are applied or removed,  TA(0.05), TR(0.05), BA(0.2) and BR(0.2) (Fig.~\ref{SNRall}a). Lowest SNR($V$), below 5, 
is observed when 0.15~g load is applied to a protoplasmic tube, and when 3~g load is applied to or removed of Physarum occupying 
agar blobs (Fig.~\ref{SNRall}, TA(0.15), BA(3), BR(3)). Highest SNRs of width, SNR($W$) over 2, are recorded during application of 
0.35~g load and removal of 0.1~g load; lowest SNR($W$) is when 0.2~g load is removed from agar blob.   

Distribution of events on the $V''/V'$ versus $W''/W'$ map (Fig.~\ref{SNRall}b) shows that for the majority of stimulations amplitude of oscillations
in a prolonged response to stimulation is 1.4 to 2.2 times higher than amplitude of oscillations of intact Physarum. Most pronounced response, 
amplitude wise, is observed when 0.15~g is lifted from a protoplasmic tube; less pronounced responses are typical for the stimulation 
events TR(0.1), TR(0.01) and BA(3).

\section{Discussion}

 We found that  slime mould of \emph{P. polycephalum} reacts to application and removal of a load 
 with a high amplitude impulse and with a temporary change of its  oscillatory activity pattern. 
There are two possible types of response: an immediate response in a form of a high-amplitude impulse, 
and, a prolonged response, in a form of changes in oscillation frequency and amplitude. We did not find a 
rigorous correlation between weight of object applied and amplitude of the 
 response impulse or amplitude and frequency of oscillations. For the time being, we can definitely claim 
 that slime mould \emph{P. polycephalum} can be used at least as  ON-OFF tactile sensor.

\begin{figure}[!tbp]
\centering
\subfigure[]{\includegraphics[width=1.1\textwidth]{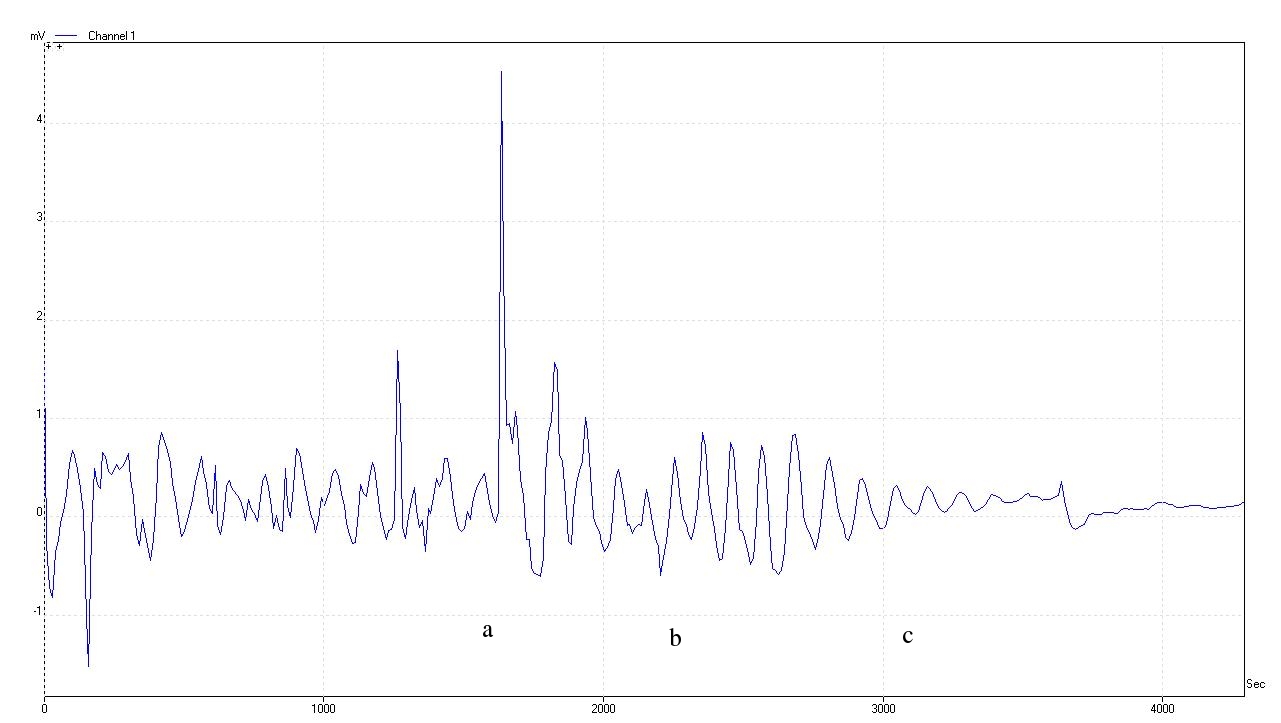}}
\caption{TA(0.05) stimulation took place at 1500~sec after beginning of experiment. 
Vertical axis is an electrical potential value in mV, horizontal axis is time in seconds.
(a)~0.05~g load is applied. 
(b)~Envelope of waves is generated. (c)~Oscillations are damped due to collapse of protoplasmic tube.}
\label{tubecrashed}
\end{figure}

What happens when an object is placed across a protoplasmic tube or on top of a fine-meshed protoplasmic network? 

When a protoplasmic tube becomes totally crashed by an applied weight (quite a rare situation) we can observe a 
typical high-amplitude impulse response. The high-amplitude response is followed by a series of enhanced oscillations and an envelope of waves.
Then damping oscillations take place indicating that the protoplasmic tube almost loses it conductivity. 
Such scenario is illustrated in Fig.~\ref{tubecrashed}. A load is applied at c. 1500~sec, the slime mould responded with 
a c. 5~mV impulse. The impulse is followed by four oscillations decreasing in amplitude. Then at c. 2100~sec an envelop of 6-7 waves was generated.
The envelop ended at c. 2900, and then Physarum generated a classical damping oscillation pattern till the tube lost its conductivity at c. 3700.

\begin{figure}[!tbp]
\centering
\subfigure[]{\includegraphics[width=0.2\textwidth]{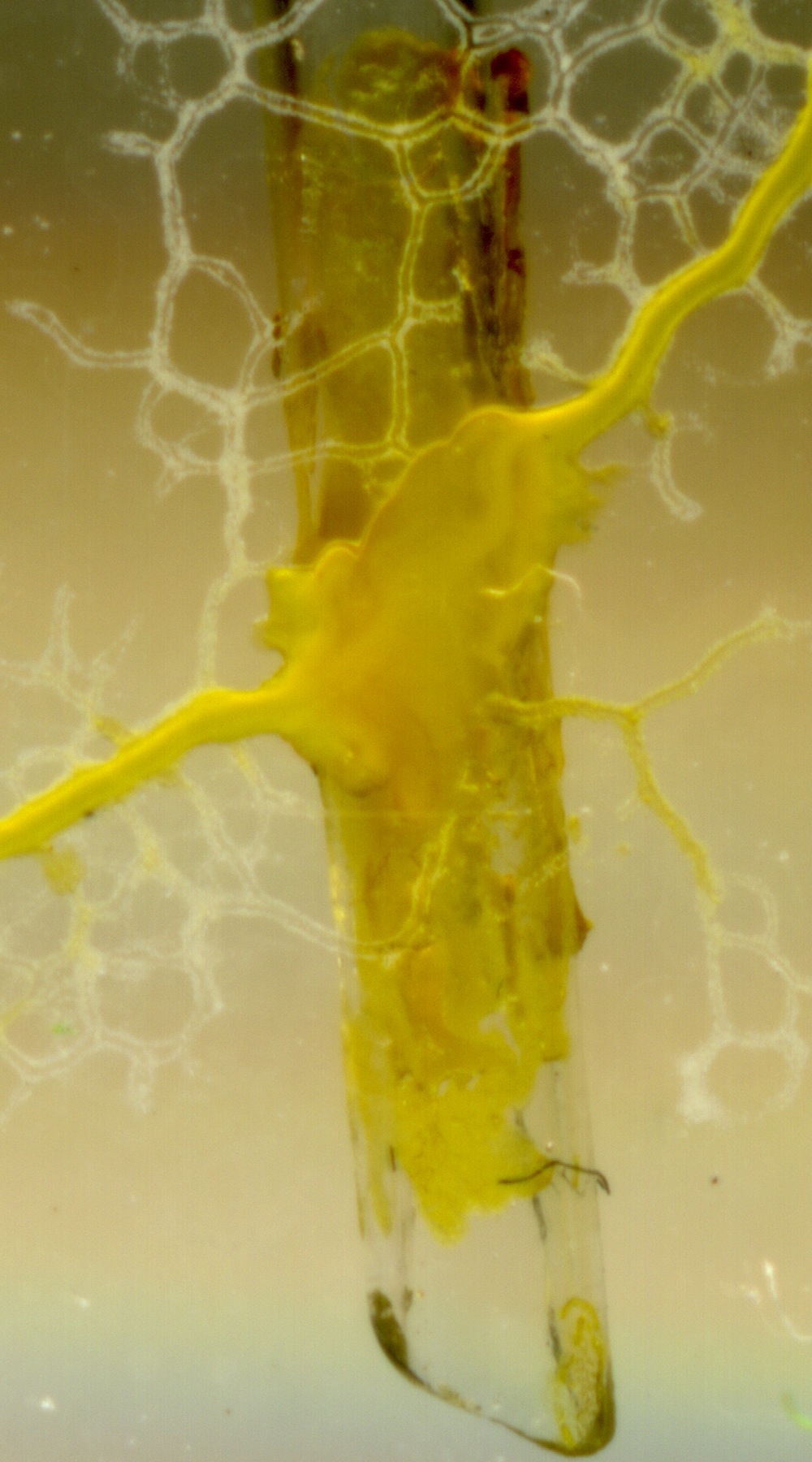}}
\subfigure[]{\includegraphics[width=0.38\textwidth]{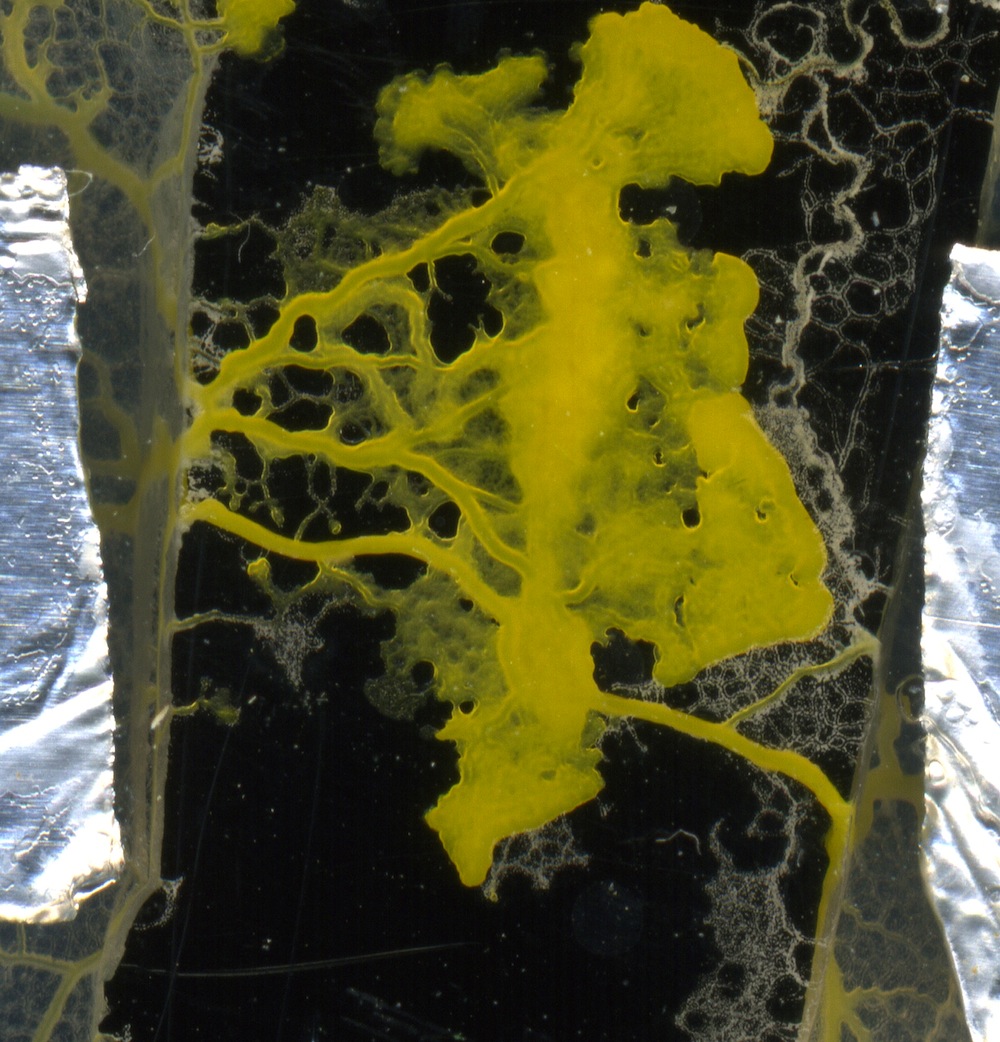}}
\subfigure[]{\includegraphics[width=0.40\textwidth]{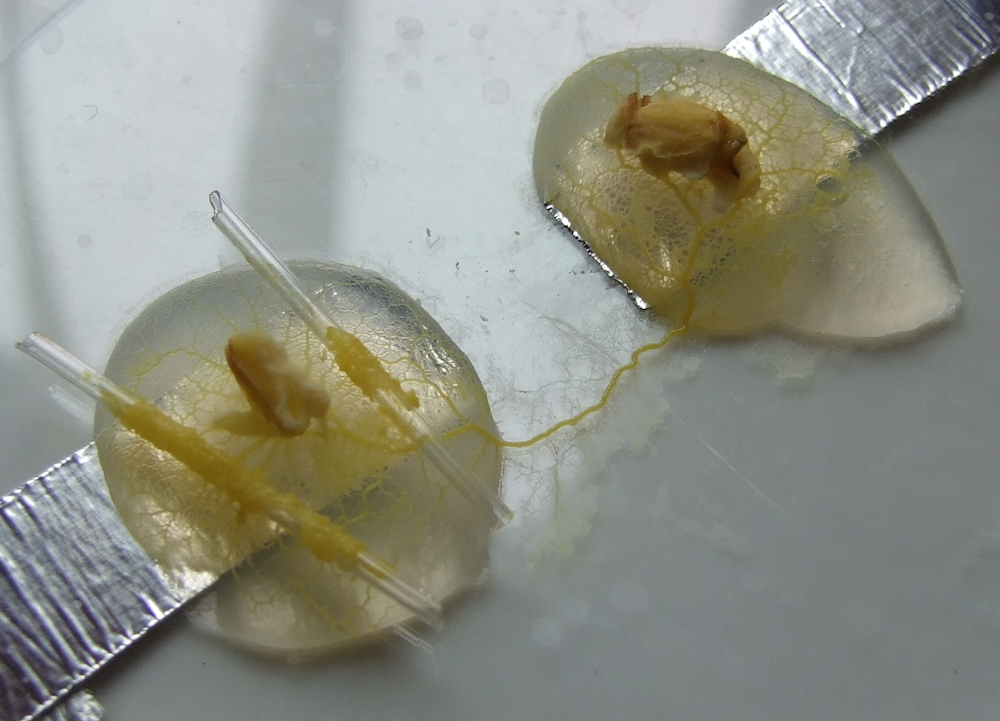}}
\subfigure[]{\includegraphics[width=0.40\textwidth]{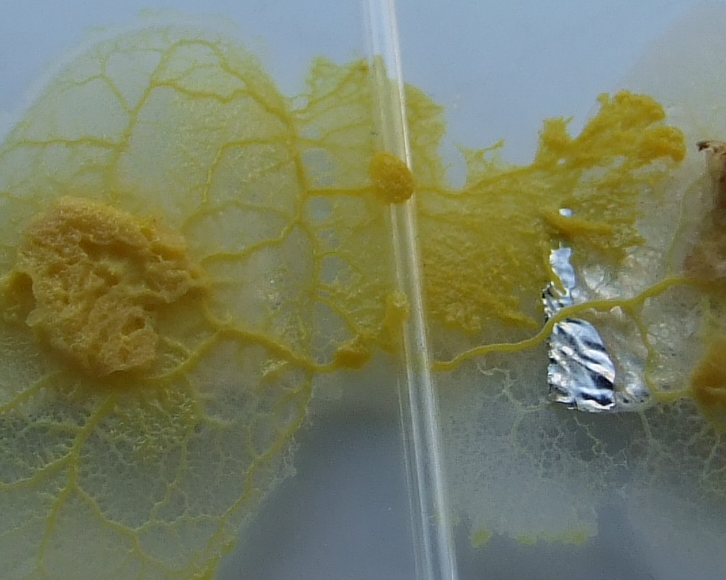}}
\subfigure[]{\includegraphics[width=0.35\textwidth]{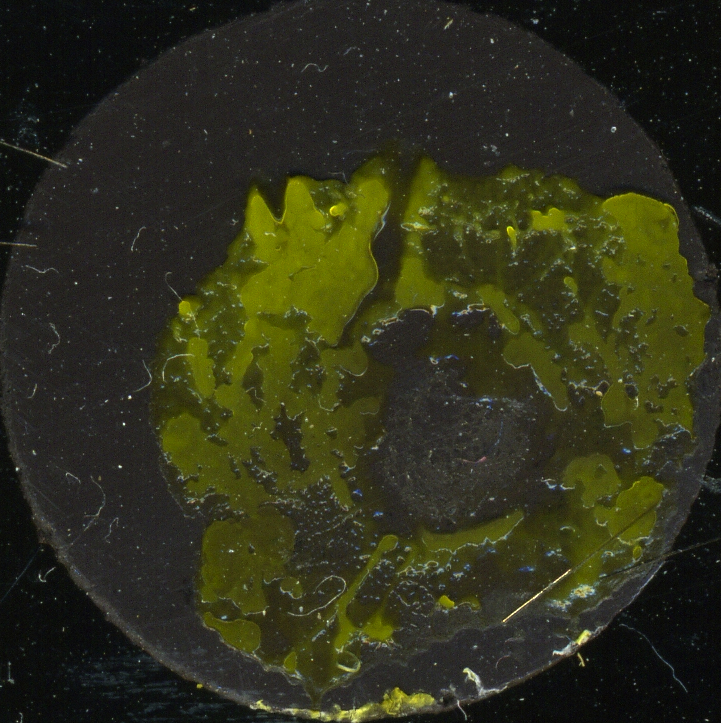}}
\caption{Physarum's morphological responses towards mechanical contact. 
(a)~A protoplasmic tube is distorted by a glass capillary placed across the tube. 
(b)~A zone of extensive growth of Physarum under and at the edges of the glass capillary.
(c)~Two segments of glass capillary placed on top of  Physarum, on agar blob, are partly colonised by the slime mould. 
(d)~Physarum colonises plastic disc placed on top of Physarum sheet wrapping agar blob, view from below.}
\label{squashedtube1}
\end{figure}

In most cases, though, the tube does not crash completely. Usually, its outer walls become deformed yet Physarum does not show any 
avoidance but instead tries to  colonisation, or encapsulation, the foreign object.  Expansion and, later, growth of a protoplasmic tube 
under a glass 
capillary is illustrated in Fig.~\ref{squashedtube1}a. If a glass capillary applied for a substantial period of time, e.g. 15-30~min, a domain 
of plasmodium matching the shape of capillary develops under and along the capillary (Fig.~\ref{squashedtube1}b).

In parallel to growing under an object Physarum crawl on the object and starts enveloping it with the plasmodial mass (Fig.~\ref{squashedtube1}cd). 
A heavy load, as e.g. plastic disc weighting 0.35g, causes collapse of the Physarum body in the immediate loci of application, however parts of Physarum closest to the 'impact zone' start colonisation of the object (Fig.~\ref{squashedtube1}e). The colonisation activities take substantial 
period and therefore unlikely to affect electrical response to mechanical stimulation. With regards to removal of a mechanical pressure, all 
objects were usually removed  c. 10~min after application, when no substantial attachment of Physarum to object occurred.

\begin{figure}[!tbp]
\centering
\includegraphics[width=1.1\textwidth]{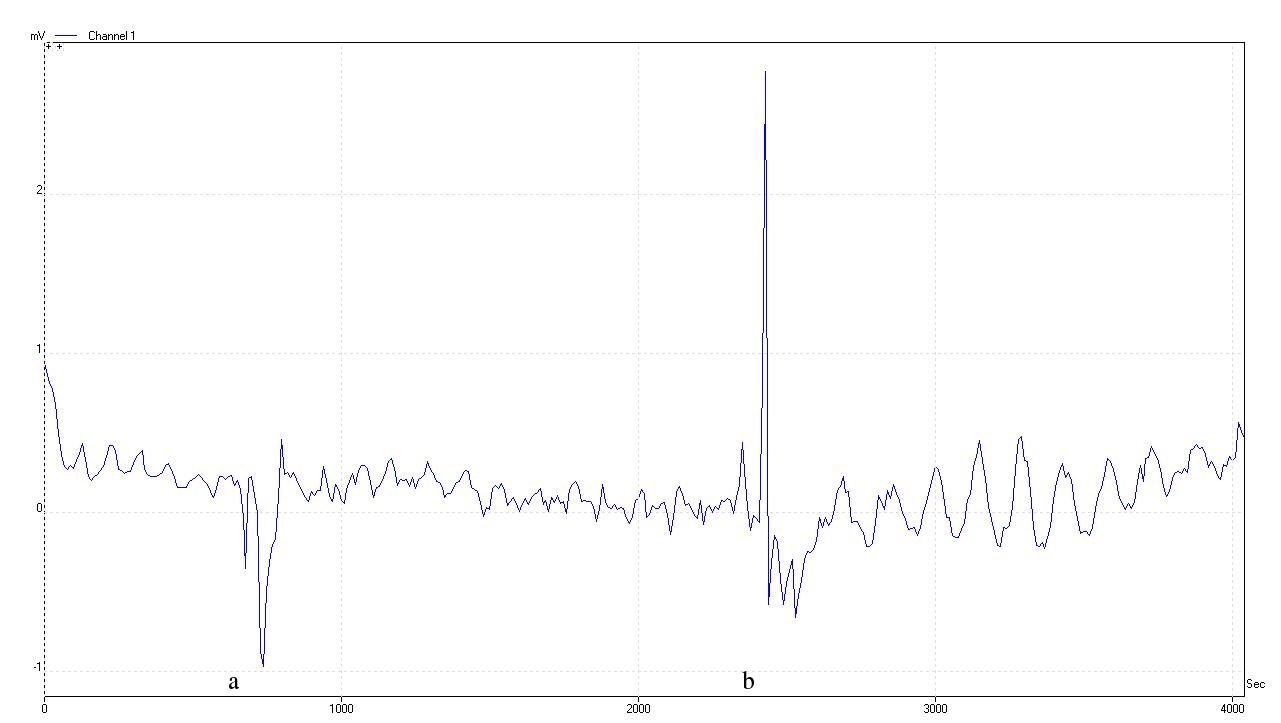}
\caption{Load, 0.2~g, is applied to Physarum no agar blob at 650~sec (a), then the agar 
blob is touched with wooden stick at 2400~sec (b), Physarum responded to both 
stimulation events with a high-amplitude impulse followed by an envelop of waves.}
\label{penciltouchA}
\end{figure}

\begin{figure}[!tbp]
\centering
\includegraphics[width=1.1\textwidth]{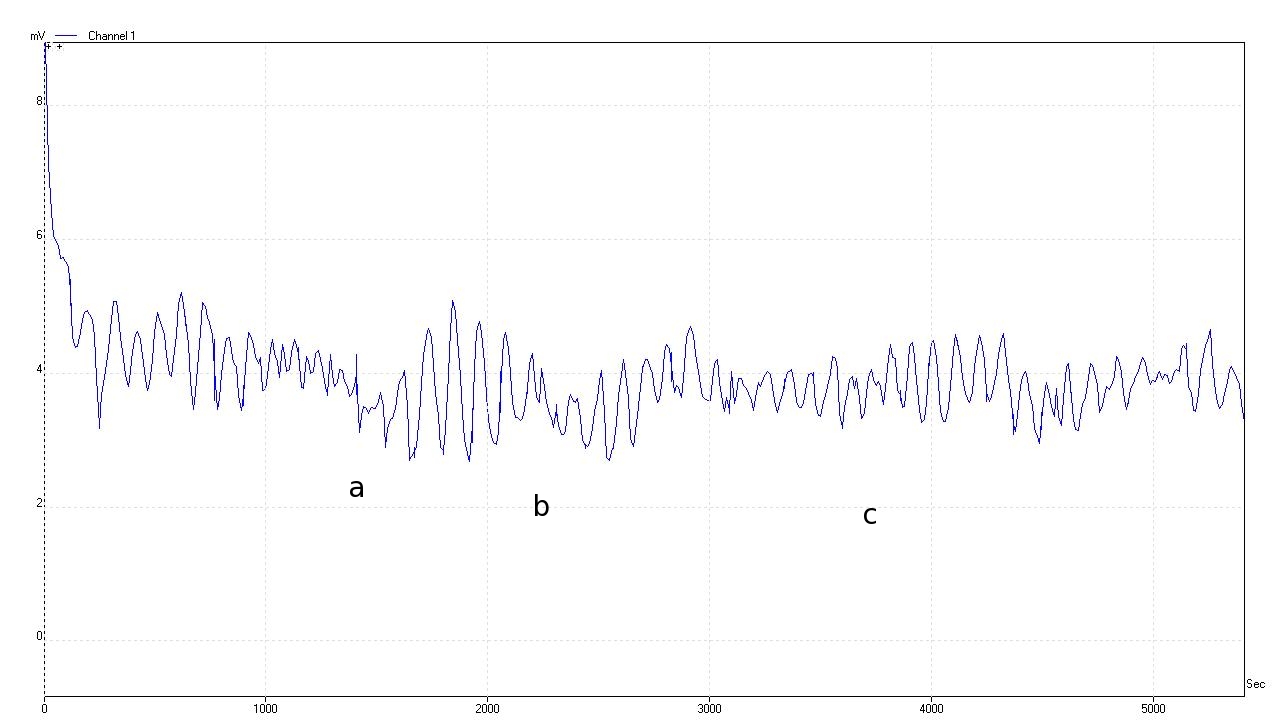}
\caption{Physarum occupying agar blob is touched with a plastic stick at 
1420~sec (a), 2230~sec (b) and 3840~sec (c) from beginning of the recording. Vertical axis is an electrical potential 
value in mV, horizontal axis is time in seconds.}
\label{penciltouchB}
\end{figure}

In the paper we systematically studied prolonged applications of loads to Physarum. 
How does Physarum react to a quick touch with a non-conductive object?  
We undertook few scoping experiments and found that apparently the slime mould exhibits similar reactions when briefly 
touched by an insulator. Two examples are shown in Figs.~\ref{penciltouchA} and \ref{penciltouchB}.  Fig.~\ref{penciltouchA} illustrates Physarum reaction to  subsequent stimulation with different tactile stimuli. First we applied 0.2~g load to a Physarum colonising an agar blob (650~sec). Physarum responded with a high-amplitude impulse and an envelop of 4-5 waves. Then we briefly touched the Physarum on agar blob with plastic stick. (2400~sec). This stimulation resulted in a very high-amplitude impulse and a delayed yet prominent envelop of  3-4 waves.

We believe an envelop of large-amplitude waves is a more typical, than a very high-amplitude impulse, 
response to a brief touch. For example, in an experiment illustrated in Fig.~\ref{penciltouchB} we touched Physarum, on agar blob, 
three times: 1420, 2230 and 3840~sec from beginning of recording. In all three cases no high-amplitude impulses were generated but
envelops of large amplitude waves were undoubtedly present. 

Physarum-based tactile sensors are certainly good in detecting dynamic forces, and possibly could even detect a slip of a load 
but they are not good in responding   to static forces. Other disadvantages of Physarum tactile sensors are relatively low, 
comparing to conventional electronic components, repeatability and susceptibility to temperature and humidity changes. 
Also sometimes a response could depend on an exact morphology of protoplasmic tubes linking the blobs and electrodes. 
Advantages of Physarum sensors are  low power consumption, almost zero costs, low design complexity, high sensitivity, 
ability to self-assemble and grow in a controllable manner. 

Our research generated more questions than given answers. Does a high-amplitude impulse bear any resemblance with solitary waves
generated by plants in a response to mechanical and thermal stimulation~\cite{land_2008}? How long an envelop of 
waves, generated in a response to stimulation, propagates without changing its state? What is a spatial resolution of a protoplasmic tube and
protoplasmic sheet slime mould sensors? What would be a dynamic of Physarum electrical response to `applied force versus time' and 
`sensing signal versus time' maps?  How Physarum will respond to a tactile stimulation in several points simulataneously?  
What is a dynamic range of Physarum tactile sensors?  How exactly the Physarum tactile sensors can be interfaced with 
and integrated into existing robotic systems?     These will be topics of further studies.

\section{Acknowledgement}

This work was supported by  Samsung Global Outreach research award in Next Generation ICs \& Interconnections.

\end{document}